\documentstyle[pre,aps]{revtex}

\begin{document}
%\draft
\preprint{MA/UC3M/03/94}

\title{Theoretical approach to two-dimensional traffic flow models}

\author{Juan M.\ Molera\thanks{E-mail: {\tt molera} or {\tt dopico} or
{\tt cuesta} {\tt @ing.uc3m.es}}, Froil\'{a}n C.\ Mart\'{\i}nez$^*$,
and Jos\'{e} A.\ Cuesta$^*$}
\address{Escuela Polit\'{e}cnica Superior, Universidad Carlos III de
Madrid, \\ c/\ Butarque 15, E-28911 Legan\'{e}s, Madrid, Spain}

\author{Ricardo Brito\thanks{E-mail: {\tt brito@seneca.fis.ucm.es}}}
\address{Facultad de Ciencias F\'{\i}sicas, Universidad Complutense,
E-28040 Madrid, Spain}

\date{\today}

\maketitle

\begin{abstract}

In this paper we present a theoretical analysis of a recently proposed
two-dimensional Cellular Automata model for traffic flow in cities with
the novel ingredient of turning capability. Numerical simulations of
this model show that there is a transition between a freely moving
phase with high velocity to a jammed state with low velocity. We study
the dynamics of such a model starting with the microscopic evolution
equation, which will serve as a basis for further analysis. It is
shown that a kinetic approach, based on the Boltzmann assumption, is
able to provide a reasonably good description of the jamming
transition. We further introduce a space-time continuous
phenomenological model leading to a couple of partial differential
equations whose preliminary results agree rather well with the
numerical simulations.

\end{abstract}

\vspace{1cm}


%\narrowtext
\section{Introduction}
\label{sec:introduction}

Traffic flow in networks is a problem of great technological
interest. In systems where good communications are crucial (cities,
computers, computer networks, etc ...) the understanding of how
undesirable, though possible, ``traffic" jams form and evolve can
lead to significant advances in design and planning of optimal
strategies \cite{Libros}.

Traffic flow is a complex collective phenomenon that presents a rich
phenomenology and, for the same reason, is difficult to deal with,
both theoretical and numerically. In spite of that, given its
great practical and technological impact, traffic car behavior in
roads has been extensively studied. Fluid dynamics and kinetic
theory have been traditionally used \cite{oldfluid} as the theoretical
tools to deal with the problem and treatment has been focused mainly on
one dimensional situations (highways with one or several lanes, with,
maybe, some intersections, is the typical example). Only recently
\cite{first_ca} models based on cellular automata (CA) have been
proposed to study this kind of problems. They present the advantage
of being easier to handle numerically while catching the correct
behavior of fluid type systems. In the last years extensive studies
on different one dimensional CA models have been carried out showing
good agreement with previous theoretical and experimental results.
These studies have also revealed new features of traffic flow. It
has been claimed, for example, the existence of
self-organized criticality in the behavior of car traffic in
highways \cite{soc}. The situation is different in two dimensions (2D). The
problem has not been treated in the fluid dynamics framework and CA
models have opened the possibility to study them.

To our knowledge only two 2D models \cite{Bihametal,nos1} have been
reported. Both of them present a periodic city with two populations of
cars moving on it. Ref.\ \cite{Bihametal} is a deterministic model
where cars never turn while in the models in Ref.\ \cite{nos1} are
stochastic and cars are allowed to turn with certain probability.
Extensive numerical simulations on them \cite{Bihametal}--\cite{naga}
show the presence of a phase transition between a free moving (high
velocity) uniform state and a jammed (low velocity) state, where the
different populations of cars are separated. In the light of recent
studies on CA models \cite{separa} the arising of this separation can
be understood as a consequence of the violation of the semidetailed
balance, characteristic of traffic models.

At the moment there is no theoretical basis to explain all these
results and it seems neccesary to build one. The only exception has
been a simple mean-field like study \cite{naga_th} of the
jamming trasition of an extension of the model in Ref.\
\cite{Bihametal}. In this paper we are going to perform a more
ambitious theoretical analysis on which further work could be based.
We will refer, as our working
model, to the one called model A in Ref.\ \cite{nos1} (briefly
described in \ref{sec:model}), but our procedure can be easily
extended to other 2D CA models.

We will begin by writing down the microscopic evolution equations for
a set of boolean variables that describe exactly the dynamics of the
system (section \ref{sec:micros}). These equations should be taken as
the starting point for any further approximation. Using the
microscopic variables we will be able to give an exact expression for
the mean velocity of the system. We will show in section
\ref{sec:lowdensity}, by means of very simple arguments, how the
behavior of the mean velocity is correctly predicted in the freely
moving phase. In section \ref{sec:Boltzmann} we
will use the tools of Kinetic Theory to obtain an approximate
macroscopic version of the microscopic equations by introducing an
average over a non-equilibrium ensemble. We have started by
considering the simplest approach: the Boltzmann approximation which
assumes the breaking of spatial two-point correlations. Though simple
this approximation already shows the presence of a transition (by
predicting the unstability of the uniform phase) and the correct
behavior of the velocity before and after the transition. It fails
however in giving the density at which the transitions occurs.
Finally, in section \ref{sec:phenomeno}, we present a continuous (in
space and time) description of the model. The purpose of this
approach is to make the connection with the previous knowledge on
traffic flow coming from fluid mechanics. The novelty is that, as
far as we know, this is the first time that a two component fluid
model has been presented to study 2D traffic flow. The continuous
approximation will lead to a system of two coupled nonlinear partial
differential equations that govern the evolution of the densities of
the two types of cars present in the city. In fact they correspond to
some limit of the Boltzmann equations of section \ref{sec:Boltzmann}.
We will see how these equations also predict a phase transition as well
as they give the correct behavior of the velocity of the two phases. We
will end up by writing our conclusions in section \ref{sec:conclusions}.

\section{Model}
\label{sec:model}

As announced in the Introduction,
in this paper we are going to deal with model A of Ref.\
\cite{nos1}. Briefly, this model can be described as follows.
We have cars moving inside a town. The town is made of
one-way perpendicular ($L$ horizontal and $L$ vertical)
streets arranged in a square lattice with periodic boundary
conditions. Vertical streets are oriented upwards and horizontal
streets are oriented rightwards. There are $N$ cars moving inside this
city. Cars sit at the crossings, and they can move to one of its
nearest neighbors (allowed by the direction of the streets) every time
step. Two cars cannot be simultaneously at the same crossing. Among
the cars, $N/2$ of them move rightward with probability $\gamma$
and upward with probability $1-\gamma$ (for symmetry reasons
$0\leq\gamma\leq 1/2$), and the remaining
$N/2$ do it the other way around. Accordingly, half of the cars move
mainly upward and half mainly rightward. Finally, there are traffic
lights that permit vertical motion at even time steps and horizontal
motion at odd time steps.

The dynamics of the model is as follows. Every time step the direction
where to move next is decided for every car according to its assigned
probability. Then, it is checked that the site where it is going to
move is empty and that the motion is allowed by the traffic light;
otherwise the car will not be moved. Finally,
all cars that can be moved are placed at their destination site and the
next time step starts. We want to stress that the whole process is
carried out simultaneously for all cars, what implies that if a car
leaves a site empty at a given time step, $t$, it will not be occupied
at the same time step. The fact that traffic lights
allow motion alternatively in vertical and horizontal streets prevents
two cars from colliding at any crossing.

Notice finally that by setting $\gamma=0$ we recover the deterministic
model of Ref.\ \cite{Bihametal}.

The main results of this model are summarized in Fig.\ \ref{phase},
which represents a plot of the average velocity ($v$) vs.\ the
density of cars ($n$) for different values of $\gamma$. The most
striking feature of this figure is the first-order-like transition
from a phase in which cars move freely to another in which cars are
jammed. We refer to Ref. \cite{nos1} for more details. In
this reference it was shown that in the jammed phase cars arrange in
diagonal strips -- with the two types of cars roughly separated in two
halves, and the question
arose whether it was one or more strips. Arguments were given supporting
the idea that, in very large systems, the jammed phase has got a
multistrip structure (in other words, the strips of this phase seem to
have a ``characteristic'' size). In this paper we will try to reproduce
analytically the phase diagram of Fig.\ \ref{phase} and to
provide an answer to what the structure of the jammed phase is like.

\section{Microscopic equations}
\label{sec:micros}

The model can be described by a set of boolean
variables for every site
of the lattice and every time step. This variables are the following:
\begin{itemize}
\item $\mu_{\bf r}^t$, the occupation number of site ${\bf r}$ at time
$t$ by a car with probability $\gamma$ of moving horizontally (1 if
occupied, 0 if not);
\item $\nu_{\bf r}^t$, the occupation number of site ${\bf r}$ at time
$t$ by a car with probability $\gamma$ of moving vertically;
\item $\xi_{\bf r}^t$, a random variable which is 1 with probability
$\gamma$ and 0 with probability $1-\gamma$;
\item $\eta_{\bf r}^t$, a random variable which is 0 with probability
$\gamma$ and 1 with probability $1-\gamma$;
\item $\sigma^t = t\, {\rm mod}\, 2$, which is 1 if traffic lights
permit horizontal movements and 0 if they permit vertical movements.
\end{itemize}
A given state of the system is completely characterized by specifying
the whole set of occupation numbers, with the constraint $\mu_{\bf r}^t
\nu_{\bf r}^t=0$ (no two cars simultaneously at the same site). The
random variables $\xi_{\bf r}^t$, $\eta_{\bf r}^t$,
$\xi_{{\bf r}'}^{t'}$ and $\eta_{{\bf r}'}^{t'}$) are all independent
two by two if ${\bf r}\ne{\bf r}'$
or $t\ne t'$; furthermore, they are also independent of the occupation
numbers and of $\sigma^t$. These random variables are introduced
to decide where cars are going
to move to: $\xi_{\bf r}^t=1$ ($\eta_{\bf r}^t=1$) decides whether a
$\mu$($\nu$)-type car placed at ${\bf r}$ at time $t$ is going to move
horizontally and $\xi_{\bf r}^t=0$ ($\eta_{\bf r}^t=0$) vertically.

With this set of boolean variables, the evolution equations for the
occupation numbers turn out to be:
\begin{equation}
\begin{array}{rccccl}
\mu_{\bf r}^{t+1} & = & \mu_{\bf r}^t & \bigl\{ &
          \sigma^t\overline{\xi}_{\bf r}^t+
          \sigma^t\xi_{\bf r}^t\left(\mu_{{\bf r}+{\bf x}}^t+
                        \nu_{{\bf r}+{\bf x}}^t \right)  \\
 & & & + & \overline{\sigma}^t\xi_{\bf r}^t+
           \overline{\sigma}^t\overline{\xi}_{\bf r}^t
                  \left(\mu_{{\bf r}+{\bf y}}^t+
                        \nu_{{\bf r}+{\bf y}}^t\right) & \bigr\} \\
 & + & \overline{\mu}_{\bf r}^t\overline{\nu}_{\bf r}^t & \bigl\{ &
           \sigma^t\xi_{{\bf r}-{\bf x}}^t\mu_{{\bf r}-{\bf x}}^t
         + \overline{\sigma}^t\overline{\xi}_{{\bf r}-{\bf y}}^t
           \mu_{{\bf r}-{\bf y}}^t  & \bigr\}  \, ,
\end{array}
\label{eveq}
\end{equation}
where we have introduced the short-hand notation $\overline{b}=1-b$ and
where ${\bf x}$ and ${\bf y}$ are vectors representing the displacement
of one site to the right and up respectively. The evolution equation for
$\nu_{\bf r}^t$ can be obtained from (\ref{eveq}) by simply exchanging
$\mu$ and $\nu$, and $\xi$ and $\eta$ (in what follows, whenever we
write down an equation for the $\mu$-type cars, it must be understood
that there is also another one for the $\nu$-type cars, which can be
obtained by this or a similar replacement). In this way, the complete
evolution of the system is specified by the above set of $2L^2$ boolean
equations plus a given initial state $\{\mu_{\bf r}^0,
\nu_{\bf r}^0\}$ and periodic boundary conditions.

Equation (\ref{eveq}) can be obtained from simple considerations on
boolean variables. First of all, each boolean variable represents a
condition that may or may not hold (for instance, $\mu_{\rm r}^t$ can
be read as ``if there is a $\mu$-type car at site ${\bf r}$ and time
$t$...'', and so on); secondly, the product of boolean variables
corresponds to the logical AND operation (whose result is always a
new boolean variable), and the addition of boolean variables turns out
to be the logical OR operation (whose result is another boolean
variable, provided all terms are mutually exclusive, i.e. of zero
product). Now the purpose is to build a boolean variable which is 1 if
there is a $\mu$-type car at site ${\bf r}$ at time $t+1$ (Eq.\
(\ref{eveq})) and 0 if there is not. This is simply achieved by
considering that (i) if
there is a $\mu$-type car at site ${\bf r}$ and time $t$ there will
still be there if it cannot move to the randomly chosen neighbouring
site (${\bf r}+{\bf x}$ or ${\bf r}+{\bf y}$), either
because it is occupied by any car or because the traffic light is
red, and (ii) if there is no car at site ${\bf r}$ and time $t$ there
will be one $\mu$-type car the next time step if there is one of those
cars at any of the neighbouring sites (${\bf r}-{\bf x}$ or
${\bf r}-{\bf y}$) willing to move to site ${\bf r}$ and its traffic
light allows the movement.
This provides all the terms of Eq.\ (\ref{eveq}); for instance,
$\mu_{\bf r}^t\sigma^t\overline{\xi}_{\bf r}^t$ is 1 only if there is a
$\mu$-type car at site ${\rm r}$ at time $t$, the traffic lights allow
horizontal movement but it decides to move vertically; $\mu_{\bf r}^t
\sigma^t\xi_{\bf r}^t(\mu_{{\bf r}+{\bf x}}^t+\nu_{{\bf r}+{\bf x}}^t)$
is 1 only if there is a $\mu$-type car at site ${\rm r}$ at time $t$,
the traffic lights allow horizontal movement, it decides to move
horizontally, but the site ${\bf r}+{\bf x}$ is occupied by any car;
etc. The meaning of the remaining terms can be interpreted in the same
way. It can be straightforwarly checked that all the terms are mutually
exclusive (remember that $b\overline{b}=0$) and that all possibilities
have been considered.

Now, we can make use of the constraint $\mu_{\bf r}^t\nu_{\bf r}^t=0$,
which lets us write $\overline{\mu}_{\bf r}^t\overline{\nu}_{\bf
r}^t=1-\mu_{\bf r}^t-\nu_{\bf r}^t$, to transform Eq.\ (\ref{eveq})
into
\begin{eqnarray}
\mu_{\bf r}^{t+1} & = &
(\sigma^t\overline{\xi}_{\bf r}^t + \overline{\sigma}^t\xi_{\bf r}^t)
\mu_{\bf r}^t
+\sigma^t\xi_{{\bf r}-{\bf x}}^t\mu_{{\bf r}-{\bf x}}^t
+\overline{\sigma}^t\overline{\xi}_{{\bf r}-{\bf y}}^t
\mu_{{\bf r}-{\bf y}}^t   \nonumber   \\
 & + &
 \sigma^t\xi_{\bf r}^t\mu_{\bf r}^t
(\mu_{{\bf r}+{\bf x}}^t+\nu_{{\bf r}+{\bf x}}^t)
-\sigma^t\xi_{{\bf r}-{\bf x}}^t\mu_{{\bf r}-{\bf x}}^t
(\mu_{\bf r}^t+\nu_{\bf r}^t)     \nonumber   \\
 & + &
 \overline{\sigma}^t\overline{\xi}_{\bf r}^t\mu_{\bf r}^t
(\mu_{{\bf r}+{\bf y}}^t+\nu_{{\bf r}+{\bf y}}^t)
-\overline{\sigma}^t\overline{\xi}_{{\bf r}-{\bf y}}^t
\mu_{{\bf r}-{\bf y}}^t(\mu_{\bf r}^t+\nu_{\bf r}^t)         \, .
\label{reeveq}
\end{eqnarray}
In this form it can be clearly seen that the highest order terms
contributing to the evolution equations are quadratic in the
occupation numbers, something
whose relevance will be made clear in the next section.

The initial configuration must be such that it satisfies the
requirements $\sum_{\bf r}\mu_{\bf r}^0=
\sum_{\bf r}\nu_{\bf r}^0=N/2$, and $\mu_{\bf r}^0\nu_{\bf r}^0=0$,
the former counts the number of cars of each type whereas the latter
express the impossibility of having two cars at the same site. It is
straightforward to check that
\begin{equation}
\sum_{\bf r}\mu_{\bf r}^t=\sum_{\bf r}\nu_{\bf r}^t=\frac{N}{2}
\label{conserv1}
\end{equation}
are constants of motion for the evolution equations (\ref{reeveq})
and that $\mu_{\bf r}^t\nu_{\bf r}^t=0$ is maintained in the evolution.

We need now a microscopic definition for the {\em average} velocity in
order to be able to obtain what could be called the `equation of
state' (average velocity vs.\ density of cars) for this model. If
we define the {\em mean} velocity at time $t$ as the number of moved
cars divided by $N$, this can be expressed in terms of the occupation
numbers:
\begin{equation}
v(t)=\frac{1}{2N}\sum_{\bf r}\left\{\left(\mu_{\bf r}^{t+1}-
\mu_{\bf r}^t\right)^2 + \left(\nu_{\bf r}^{t+1}-
\nu_{\bf r}^t\right)^2\right\}\, . \label{mvel}
\end{equation}
This equation can be easily visualized if one realizes that
$\mu_{\bf r}^{t+1}-\mu_{\bf r}^t$ is zero only if there is no
$\mu$-type car at site ${\bf r}$ at times $t$ and $t+1$ or if there is
one that remains there at both times; in other words, if there is no
movement of any $\mu$-type car at site ${\bf r}$ between $t$ and $t+1$.
Otherwise the value is 1 (if the car arrives at time $t+1$) or $-1$ (if
the car leaves off at time $t+1$). The square removes the sign and the
factor $1/2$ let the expression count only the cars that arrive and not
those which leave. The same applies to the term related to $\nu$-type
cars. Notice that only one of the two terms may be 1.

The average velocity was computed in Ref.\ \cite{nos1} as the limit
\begin{equation}
v=\lim_{T\to\infty}\frac{1}{T}\sum_{t=0}^Tv(t) \, ,
\label{avel}
\end{equation}
which, if the system reaches a steady state, corresponds to its mean
velocity \cite{note1}.

Before we go any further we would like to remark that all the formalism
introduced in this section also applies to the deterministic model
\cite{Bihametal} by simply setting $\gamma=0$, although in what follows
we will make explicit use of the randomness to obtain information out of
the evolution equations.

\section{Low density limit}
\label{sec:lowdensity}

Figure \ref{phase} shows that the behaviour of the $v$-$n$ curves in the
free moving phase is, to a large accuracy, $\gamma$-independent (for
$\gamma>0$). In fact, they decrease almost linearly with slope
$-1/2$. Reproducing analytically this remarkable property is a first
important result which can be derived from the microscopic equations.
It will be the main goal of this section.
To this purpose notice that Eq.\ (\ref{mvel}) can be rewritten
\begin{equation}
v(t)=1-\frac{1}{n}\left(\left[\mu^{t+1}\mu^t\right]+
\left[\nu^{t+1}\nu^t\right]\right) \, ,
\label{mvel2}
\end{equation}
where the following notation has been introduced:
$\left[A^t\right]\equiv L^{-2}\sum_{\bf r}A_{\bf r}^t$, the
average over the whole lattice of a given function $A_{\bf r}^t$,
and the car density, $n$, defined as $n\equiv NL^{-2}$. In getting
(\ref{mvel2}) it has been used that any boolean variable $b$ verifies
$b^2=b$ and that $\left[\mu^t\right]=\left[\nu^t\right]=n/2$. Time
correlations can be transformed into space correlations by introducing
the dynamics, i.e., by using the equations of motion (\ref{reeveq}):
\begin{eqnarray}
\left[\mu^{t+1}\mu^t\right] & = & \sigma^t\left\{
\left[\overline{\xi}^t\mu^t\right]+
\left[\xi^t\mu^t\mu_{\bf x}^t\right]+
\left[\xi^t\mu^t\nu_{\bf x}^t\right] \right\}  \nonumber  \\
& + & \overline{\sigma}^t\left\{
\left[\xi^t\mu^t\right]+
\left[\overline{\xi}^t\mu^t\mu_{\bf y}^t\right]+
\left[\overline{\xi}^t\mu^t\nu_{\bf y}^t\right] \right\}  \, ,
\label{corrt}
\end{eqnarray}
and the counterpart for the $\nu$'s (obtained by exchanging $\xi$ and
$\eta$ and $\mu$ and $\nu$ in the equation above). We have denoted with
the subscripts ${\bf x}$ or ${\bf y}$ a translation by those unit
vectors. Equation\ (\ref{corrt}) simplifies in the $L\to\infty$ limit,
for then the random variables can be replaced by their averages
($\gamma$ or $1-\gamma$), as they are independent of the car variables.
Furthermore, in the low density limit ($n\to 0$) we can assume that the
occupation numbers become independent variables in the uniform steady
state ($\left[\mu^t\mu^t_{\bf x}\right]\sim\left[\mu^t\right]
\left[\mu^t_{\bf x}\right]=n^2/4$, and similarly for the other seven
spatial correlations -- those in (\ref{corrt}) and its counterpart.
In principle (though we will later come back to this point again) this
hypothesis is justified by the fact that, due to the random
dynamics, at very low densities cars have enough time to ``forget''
between two successive encounters. Therefore, in this regime,
\begin{equation}
\left[\mu^{t+1}\mu^t\right] \sim \sigma^t\left\{
\overline{\gamma}\frac{n}{2}+2\gamma\left(\frac{n}{2}\right)^2
\right\} + \overline{\sigma}^t\left\{ \gamma\frac{n}{2}+
2\overline{\gamma}\left(\frac{n}{2}\right)^2 \right\} \, ,
\label{corrt2}
\end{equation}
Plugging this expression into the equation for
the mean velocity (\ref{mvel2}) we get
\begin{equation}
v(t) \sim \frac{1}{2}(1-n) \, ,
\label{mvel3}
\end{equation}
and since this formula corresponds to the uniform steady state it
is the average velocity (\ref{avel}). This expression for
$v$ is $\gamma$-independent, which is in good agreement with the
results obtained in the simulations in the freely moving phase; in
particular it predicts the slope $-1/2$ of the $v$-$n$ curves for small
$n$.

Clearly, the above argument does not hold for the deterministic case
($\gamma=0$) reported in Ref.\ \cite{Bihametal}, because in the
derivation of Eq.\ (\ref{mvel3}) we have made explicit use of the
randomness. This is very clear when one compares with the result
obtained in that reference -- $v=1$ for the freely moving phase
and $v=0$ for the jammed phase. However it is still an open question
whether the jamming transition in this deterministic model drops to
$n=0$ as the system size increases; were this result confirmed Eq.\
(\ref{mvel3}) would also be valid in the limit $\gamma\to 0$ (remember
that its derivation assumes that the size of the system is infinite),
what would provide a beautifully closed picture of this system.

\section{Boltzmann approximation}
\label{sec:Boltzmann}

Equation (\ref{reeveq}) and its $\nu$-counterpart make feasible to
employ the methods of the standard Kinetic Theory
\cite{Resibois,Balescu} in its lattice-gas version \cite{DBE,six} in
order to obtain macroscopic properties of the system. To this purpose
let us introduce a non-equilibrium ensemble in which all initial
conditions, $\{\mu_{\bf r}^0,\nu_{\bf r}^0\}$, are equally weighted.
We will denote the average over this ensemble by $\langle\cdot\rangle$.
When we average over the boolean variables $\mu_{\bf r}^t$ and
$\nu_{\bf r}^t$, we obtain real variables -- $u({\bf r},t)$ and
$w({\bf r},t)$ respectively -- ranging from 0 to 1 which measure the
average number of initial configurations for which the site ${\bf r}$
is occupied at time $t$ by the corresponding type of car (henceforth
average occupation). Now, if we average over this ensemble the
microscopic evolution equations (\ref{reeveq}) we obtain
\begin{eqnarray}
u({\bf r},t+1) & = &
(\sigma^t\overline{\gamma} + \overline{\sigma}^t\gamma) u({\bf r},t)
+\sigma^t\gamma u({\bf r}-{\bf x},t)
+\overline{\sigma}^t\overline{\gamma} u({\bf r}-{\bf y},t)
\nonumber \\
 & + &
 \sigma^t\gamma\left(\left\langle\mu_{\bf r}^t
\mu_{{\bf r}+{\bf x}}^t\right\rangle -
\left\langle\mu_{{\bf r}-{\bf x}}^t\mu_{\bf r}^t\right\rangle +
\left\langle\mu_{\bf r}^t\nu_{{\bf r}+{\bf x}}^t\right\rangle -
\left\langle\mu_{{\bf r}-{\bf x}}^t\nu_{\bf r}^t\right\rangle\right)
\nonumber         \\
 & + &
 \overline{\sigma}^t\overline{\gamma}\left(
\left\langle\mu_{\bf r}^t\mu_{{\bf r}+{\bf y}}^t\right\rangle -
\left\langle\mu_{{\bf r}-{\bf y}}^t\mu_{\bf r}^t\right\rangle +
\left\langle\mu_{\bf r}^t\nu_{{\bf r}+{\bf y}}^t\right\rangle -
\left\langle\mu_{{\bf r}-{\bf y}}^t\nu_{\bf r}^t\right\rangle\right).
\label{aveveq}
\end{eqnarray}
In writing Eq.\ (\ref{aveveq}) we have used the fact that
$\xi_{\bf r}^t$ and $\eta_{\bf r}^t$ are uncorrelated of any
combination of occupation numbers at the same time step. These
equations are not closed due to the presence of the quadratic terms
which, on the other hand, contain the interaction of the model and
therefore cannot be ignored. Evolution equations for the averages of
these quadratic terms can be written, but cubic and quartic terms will
appear, and so on, giving rise to a hierarchy similar to the BBGKY one
\cite{Balescu}. Being able to write down a closed set of equations for
the average occupations requires an approximation. The simplest one is
to assume that average occupations at different sites are {\em always}
uncorrelated. In Kinetic Theory this
approximation is known as the Boltzmann approximation (or molecular
chaos hypothesis). For our purpose, this approximation amounts to
writing
\begin{mathletters}\begin{eqnarray}
\left\langle \mu_{\bf r}^t\mu_{\bf r'}^t\right\rangle & = &
u({\bf r},t)u({\bf r'},t)   \label{Bhypmm}   \\
\left\langle \mu_{\bf r}^t\nu_{\bf r'}^t\right\rangle & = &
u({\bf r},t)w({\bf r'},t)   \label{Bhypmn}   \\
\langle \nu_{\bf r}^t\nu_{\bf r'}^t\rangle & = &
w({\bf r},t)w({\bf r'},t)   \label{Bhypnn}
\end{eqnarray}\end{mathletters}\par\noindent
for any pair of different sites ${\bf r}$ and ${\bf r'}$ and for every
time step $t$. Physically speaking, the molecular chaos hypothesis
assumes that two colliding ``particles'' (cars in our model) have never
met before, or if they have, all their mutual influence has been lost.
This hypothesis is expected to be true in the limits of short times
(cars do have never met before) or vanishing densities (the probability
of an encounter is small enough so as to lose all information between
successive encounters). We have already made use of this fact in Sec.\
\ref{sec:lowdensity}.

A further simplification may be introduced, as it is the replacement
of the traffic-light variable, $\sigma^t$, by its time average, $1/2$.
This approximation is expected to hold in the macroscopic regime (which
we are interested in), where the microscopic details are not seen.
However, the study of microscopic scales would require to keep
$\sigma^t$ to its original values 0, 1 alternatively. We will go back
to this point later.

In summary, the ensemble average together with these two approximations
lead to a closed set of equations (henceforth referred to as the
{\em Boltzmann equations}) for the averaged variables $u({\bf r},t)$
and $w({\bf r},t)$:
\begin{eqnarray}
u({\bf r},t+1) & = & \frac{1}{2}u({\bf r},t) +
\frac{\gamma}{2} u({\bf r}-{\bf x},t)+
\frac{\overline{\gamma}}{2} u({\bf r}-{\bf y},t)    \nonumber \\
 & + &
\frac{\gamma}{2}u({\bf r},t)\{ u({\bf r}+{\bf x},t)
+ w({\bf r}+{\bf x},t) \} -
\frac{\gamma}{2}u({\bf r}-{\bf x},t)\{
u({\bf r},t) + w({\bf r},t) \}
\nonumber  \\
 & + &
\frac{\overline{\gamma}}{2}u({\bf r},t)\{ u({\bf r}+{\bf y},t)
+ w({\bf r}+{\bf y},t) \} -
\frac{\overline{\gamma}}{2}u({\bf r}-{\bf y},t)\{
u({\bf r},t) + w({\bf r},t) \}        \label{Boltzeq}
\end{eqnarray}
Iteration of the Boltzmann equations will give a mean-field-like
evolution of our model. This evolution will lack some of the effects
caused by correlations, but it still keeps many of the interesting
features of the model. Furthermore, the mean field description allows
us to obtain analytically many physical quantities of relevant
interest. The remaining of this section is devoted to study the
Boltzmann equations. Subsection \ref{LSA} deals with the linear regime
where we determine the stability of the homogeneous solution, whereas
in subsection \ref{BS} we present the results of a numerical
simulation performed directly on the Boltzmann equations and compare
them with the linear regime as well as with the simulations of the full
model \cite{nos1} obtained from the exact microdynamics, Eqs.\
(\ref{reeveq}).

\subsection{Linear Stability Analysis}
\label{LSA}

The Boltzmann equations (\ref{Boltzeq}) have, for any value of
the turning parameter $\gamma$, the uniform solution
\begin{equation}
u({\bf r},t)=w({\bf r},t)=n/2\, ,\quad\forall t\geq 0   \, .
\label{uniform}
\end{equation}
This is what simulations show for small enough density \cite{nos1}.
However above a threshold (depending of $\gamma$) the simulations show
a steady non-uniform pattern, what means that the uniform solution
becomes unstable against small perturbations. Thus, settling the
stability of (\ref{uniform}) requires studying the evolution of small
perturbations about it, i.e. we need to look for solutions of the form
\begin{equation}
u({\bf r},t)=\frac{n}{2}+\delta u({\bf r},t)\, ,\qquad
w({\bf r},t)=\frac{n}{2}+\delta w({\bf r},t)\, ,
\label{perturbation}
\end{equation}
which introduced into (\ref{Boltzeq}) give, to linear order:
\begin{eqnarray}
\delta u({\bf r},t+1) & = & \frac{1}{2}\left\{
\left(1+\frac{n}{2}\right) \delta u({\bf r},t) \right. \nonumber  \\
 & + & \gamma\frac{n}{2}\delta u({\bf r}+{\bf x},t) +
       \gamma(1-n)\delta u({\bf r}-{\bf x},t)   \nonumber   \\
 & + & \left.
       \overline{\gamma}\frac{n}{2}\delta u({\bf r}+{\bf y},t) +
       \overline{\gamma}(1-n)\delta u({\bf r}-{\bf y},t) \right\}
\nonumber \\
 & - & \frac{n}{4}\{ \delta w({\bf r},t) -
               \gamma\delta w({\bf r}+{\bf x},t) -
               \overline{\gamma}\delta w({\bf r}+{\bf y},t)  \} \, .
\label{expans}
\end{eqnarray}
Equation (\ref{expans}) is linear but non-local, because it involves
the average car occupations at five nodes ($\bf r$ and its four nearest
neighbours). By taking the discrete Fourier transform, defined as:
\begin{equation}
\delta\widehat{f}({\bf k},t)=\sum_{\bf r}
{\rm e}^{-i{\bf k}\cdot{\bf r}}\delta f({\bf r},t) \, ,
\end{equation}
(where $k_{x(y)}=2\pi q_{x(y)}/L$, $q_{x(y)}=0,1,2,\dots,L-1$ and $f$
is either $\mu$ or $\nu$) Eq.\ (\ref{expans}) becomes
\begin{equation}
\left( \begin{array}{c}
           \delta\widehat{u}({\bf k},t+1)    \\
           \delta\widehat{w}({\bf k},t+1)
       \end{array}
\right)
= \Omega(\bf k)
\left( \begin{array}{c}
           \delta\widehat{u}({\bf k},t)      \\
           \delta\widehat{w}({\bf k},t)
\end{array} \right) \, ,
\label{linear}
\end{equation}
where $\Omega(\bf k)$ is the ${\bf k}$-dependent 2$\times$2 linear
evolution operator whose elements are given by
\begin{mathletters}\begin{eqnarray}
\Omega_{11} & = & \frac{1}{2}+\frac{n}{4}
      \{ 1 + \gamma S_x + \overline{\gamma}S_y \} +
      \frac{1-n}{2} \{ \gamma S_x^* + \overline{\gamma} S_y^* \}
\, , \label{omega11} \\
\Omega_{12} & = & \frac{n}{4}\{\gamma S_x+\overline{\gamma}S_y-1\}
\, , \label{omega12}
\end{eqnarray}\label{omegas}\end{mathletters}\par\noindent
with $S_{x(y)}\equiv {\rm e}^{i{\bf k}_{x(y)}}$. The elements
$\Omega_{21}$ and $\Omega_{22}$ are obtained from Eqs.\ (\ref{omegas})
by exchanging $\gamma$ and $\overline{\gamma}$ in $\Omega_{12}$ and
$\Omega_{11}$ respectively. Iteration of Eq.\ (\ref{linear}) yields
the dynamics of the system at the linear level in Boltzmann
approximation. The information on the stability of the uniform state
is contained in the eigenvalues of the linear evolution operator
$\Omega({\bf k})$, written for convenience as
\begin{equation}
\Omega({\bf k})\psi_j({\bf k})={\rm e}^{z_j({\bf k})}
\psi_j({\bf k}),\qquad j=1,2  \, .
\label{eigen2}
\end{equation}
Hence the dynamics at the linear level is simply expressed as
\begin{equation}
\left( \begin{array}{c} \delta\widehat{u}({\bf k},t)     \\
\delta\widehat{w}({\bf k},t)  \end{array} \right)
= U({\bf k}) \left( \begin{array}{cc} {\rm e}^{z_1({\bf k})t} & 0 \\
0 & {\rm e}^{z_2({\bf k})t} \end{array}\right)  U^{-1}({\bf k})
\left(\begin{array}{c} \delta\widehat{u}({\bf k},0) \\
\delta\widehat{w}({\bf k},0)  \end{array} \right)  \, ,
\label{eigendynamics}
\end{equation}
where the first (second) column of $U({\bf k})$ is the eigenvector
$\psi_{1(2)}({\bf k})$. Note here that the two eigenvalues must
satisfy
\begin{equation}
\lim_{{\bf k}\to 0} z_j({\bf k})=0
\label{k0relation}
\end{equation}
as a consequence of the conservation laws (\ref{conserv1}) (for the
sake of simplicity we will be referring to $z_j({\bf k})$ as
the eigevalues, though they are actually their logarithm).

Whenever ${\rm Re}\; z_j({\bf k})<0$ ($j=1,2$) for all ${\bf k}$, the
amplitude of all eigenmodes, $\psi_j({\bf k})$, will vanish
exponentially with time and finally disappear, recovering the uniform
state. However, as soon as a value of ${\bf k}$ appears such that
${\rm Re}\; z_{j_m}({\bf k})>0$ (for $j_m=1$ or 2, or both), then the
amplitude of the corresponding eigenmode $\psi_{j_m}({\bf k})$ will
grow exponentially with time, in other words, the uniform state will
become unstable, and inhomogeneous spatial structure will develop.
Therefore, the spectrum of the linear evolution operator determines the
macroscopic behavior of the system. There is more information that can
be extracted from the spectrum when the uniform state is unstable
\cite{bussemaker}. In general, there will be a wavevector, ${\bf k}_m$,
for which ${\rm Re}\; z_{j_m}({\bf k}_m)$ (for some $j_m=1,2$) is maximum.
Fluctuations with this wavevector will dominate as $t\to\infty$ and
consequently, the structure will possess a typical lengthscale of
$\lambda_m\equiv 2\pi/k_m$, and will be formed along the direction of
${\bf k}_m$. Furthermore, the onset time of the instability is
given by $\left[{\rm Re}\; z_{j_m}({\bf k}_m)\right]^{-1}$ (see Eq.\
(\ref{eigendynamics})).

The eigenvalues $z_j({\bf k})$ are easily obtained from the elements
of the linear evolution operator, given in (\ref{omegas}). A typical
plot of ${\rm Re}\; z_{j_m}({\bf k})$ is shown in Fig.\
\ref{eigenrho=0.8p=0.1} for $n=0.8$ and $\gamma=0.1$. According to
the numerical simulations (Fig.\ \ref{phase}) for this parameter set
the system is jammed, i.e., the system is non-uniform since cars
cluster in bands which form an angle $\theta=\pi/4$ with the positive
$x$-axis. From Fig.\ \ref{eigenrho=0.8p=0.1} we see that the linear
stability analysis agrees with this fact as there exists regions where
${\rm Re}\; z_{j_m}({\bf k})$ is positive with maxima located on a line
forming an angle of $\theta=3\pi/4$ with the positive $x-$axis,
i.e., perpendicular to the bands of the system. From the location
of the maxima we can obtain the typical size of the bands or,
equivalently, the number of bands which would appear in a finite size
simulation. We will give more details about this point later in this
section. We have not shown the second eigenvalue, because it satisfies
${\rm Re}\; z({\bf k})<0$ for all values of ${\bf k}$, so it does
not affect the previous discussion.

The traffic jams of this model are always bands as those of Fig.\
\ref{phase}b (i.e. with $\theta=\pi/4$). Hence, the maxima of
${\rm Re}\; z_{i_m}({\bf k})$ appear along the direction given by
${\bf\widehat{k}}=(-1/\sqrt{2},1/\sqrt{2})$. Therefore, in order to
look for these maxima in the expressions for $z_i({\bf k})$ is enough
to choose a wavevector of the form ${\bf k}=k{\bf\widehat{k}}$. This
means that $S_x=S_y^*=\exp(-ik/\sqrt{2})$. For this particular
direction the eigenvalues take a simple form:
\begin{eqnarray}
e^{z_{\pm}(k)} & = & 1-s^2\left(1-\frac{n}{2}\right) \nonumber \\
 & \pm & s\left\{\frac{s^2n^2}{4}-2(1-n)
\left(\frac{1}{2}-n\right)\zeta^2(1-s^2)\right\}^{1/2} \, ,
\label{joseigen}
\end{eqnarray}
where $s\equiv\sin(k/2\sqrt2)$ and $\zeta\equiv 1-2\gamma$. It is
straightforward to check that ${\rm Re}\; z_+(k)>1$ for some $k$'s if
and only if $n>1/2$. In consequence, this analysis predicts a jamming
transition for $n>1/2$ irrespective the value of $\gamma$. The
existence of this transition is the most important result of this
section; furthermore, the prediction for the transition density is
reasonably good for $\gamma$ close to $1/2$, the highest randomness
(see Fig.\ \ref{phase}a). However it disagrees with the simulations for
smaller $\gamma$'s.

A comment about the role of the traffic light variables, $\sigma^t$ is
needed here. Figure \ref{eigenrho=0.8p=0.1} (and the discussion about
it) has been obtained for $\sigma^t=1/2$. Keeping the variables
$\sigma^t$ to their original values, 0, 1, alternatively, one obtains
two evolution operators, one for even time steps, $\Omega^e({\bf k})$,
and another one for odd time steps, $\Omega^o({\bf k})$. Then the
eigenvalue problem in Eq.\ (\ref{eigen2}) has to be formulated in terms
of an effective evolution operator given by $\{\Omega^e({\bf k})
\Omega^o({\bf k})\}^{1/2}$. However, in doing so, one has to face the
fact that $\Omega^o({\bf k})$ and $\Omega^e({\bf k})$ do not commute.
Thus we end up with three sets of eigenvalues, those of Eq.\
(\ref{eigen2}), of $\{\Omega^e({\bf k})\Omega^o({\bf k})\}^{1/2}$
and of $\{\Omega^o({\bf k})\Omega^e({\bf k})\}^{1/2}$. The
largest difference between these three sets is for the wavevectors
${\bf k}$ located at the edges of the Brillouin zone ($k\simeq\pi$),
related to the microscopic behaviour, when the system is observed at
scales of the order of a few lattice spacings. On the contrary, for
wavevectors at the center of the Brillouin zone ($k\simeq 0$)
the three sets coincide. This is the region we are interested in,
because it describes the macroscopic behaviour. As a
summary, the macroscopic behavior is not affected by the traffic lights
being 0,1 or its average 1/2. For this reason, and for the sake of
simplicity, we have made all the calculations with $\sigma^t$ replaced
by 1/2.

According to Kinetic Theory, for each quantity conserved by the
dynamics there exists an eigenvector as given by Eq.\ (\ref{eigen2})
whose eigenvalue satisfies the relation
(\ref{k0relation}). In the long wavelength (small wavenumber) limit,
the eigenvalues can be written as:
\begin{equation}
z_j({\bf k})= ikc_j +(ik)^2 D_j +\cdots \qquad ({\bf k}\to 0),
\label{expansion}
\end{equation}
The quantity $c_j$ is interpreted as the speed of propagation of the
mode, while $D_j$ is the corresponding diffusivity. In the case of
fluid dynamics in three dimensions, there are five conserved
quantities, and five of such modes: two sound modes, two shear modes,
and finally a thermal mode. This is a general result of Kinetic Theory
and also holds for lattice gases as models for thermal fluids
\cite{GBBE}. In the case of a model of concentration diffusion leading
to the Fick law, there is only one mode (the concentration); $c_j$ is
then the drift velocity and $D_j$ the diffussion coefficient.
The two eigenvalues of our model (Eq.\ (\ref{joseigen})) correspond to
the two conservation laws (\ref{conserv1}), so that expanding them in
powers of the wavevector $k$, as in Eq.\ (\ref{expansion}), we can
identify the two (opposite) speeds of propagation and the (unique)
diffusion coefficient, with the result (for $n<1/2$):
\begin{mathletters}\begin{eqnarray}
c_{\pm} & = & \pm\frac{1}{2}\zeta (1-n)^{1/2}
\left(n-\frac{1}{2}\right)^{1/2}         \\
D & = & \frac{1}{16}\left[2-n-(1-n)(1-2n)\zeta^2\right]
\label{vD}
\end{eqnarray}\end{mathletters}\par\noindent
We have to state here that this velocity has nothing to do with the
remnant velocity measured from the simulations, but it is the velocity
with which long-wavelenght fluctuations propagate (the so-called
kinematic velocity in fluid-dynamics studies of traffic flow
\cite{oldfluid}).

As it was said before, by calculating the maximum with respect to $k$
of ${\rm Re}\; z_+({\bf k})$, with ${\bf k}$ along the direction of
$\theta=3\pi/4$ (Eq.\ (\ref{joseigen})) we obtain the characteristic
length of the system as $\lambda_m\equiv 2\pi/k_m$. The
relevance of this quantity is that it provides information about the
typical center-to-center band separation in the jammed phase and
consequently determines the number of bands. We will not write here the
expression for $\lambda_m$, because it is very cumbersome. Instead, in
Fig.\ \ref{figlam} we plot the value of $\lambda_m$ as a function of
the density $n$ for some values of the randomness $\gamma$.
It can be seen that apart from both edges ($n\gtrsim 1/2$ and
$n\lesssim 1$), the bands have a typical width of few tens of lattice
spacings, in qualitative agreement with the simulations. However, in
actual simulations the characteristic length may not be the natural one
(that the system would have if it were infinite) because of a
competition with the finite size of the system. If the system is not
large enough to form an integer number of bands of width $\lambda_m$ it
will form less
bands, but concentrating all the cars in the existing bands. Therefore
the characteristic length will be modified. However, the value of
$\lambda_m$ we have obtained can still give a good prediction of the
most stable number of bands in the simulations at finite $L$. The
number of bands predicted by the Boltzmann theory is simply
$L/\lambda_m$. If $L/\lambda_m$ is not an integer, the system will have
the trend to form $\lfloor L/\lambda_m\rfloor$ bands (where $\lfloor
x\rfloor$ means the largest integer smaller or equal to $x$). Reading
off the  Fig.\ \ref{figlam} for the case $\gamma=0.4$ we obtain
$\lfloor L/\lambda_m\rfloor=1$ for all $n>1/2$, in perfect agreement
with what is shown Fig.\ \ref{phase}a (made for a city with
$L=64$). The absence of jumps in this curve (in the jammed phase) means
that only one band has been formed. For the case $\gamma=0.3$, Fig.\
\ref{figlam} predicts $\lfloor L/\lambda_m\rfloor=1$ for $n\lesssim
0.6$, $\lfloor L/\lambda_m\rfloor=2$ for $0.6\lesssim n\lesssim 0.7$
and $\lfloor L/\lambda_m\rfloor=3$ for $0.7\lesssim n\lesssim 0.95$.
Inspection of Fig.\ \ref{phase} shows that there is only one band for
$n\lesssim 0.65$, while for $n\gtrsim
0.65$ there are two bands, as indicated from the increase of the
velocity. For higher densities, although it cannot be seen in Fig.\
\ref{phase}, occasionally some simulations end up in a three-band
structure \cite{nos1}. Unfortunaltely, the agreement at this level
between the Boltzmann predictions and the simulations is poorer the
smaller the randomness. For $\gamma=0.2$ and $\gamma=0.1$ the number of
bands given by the Boltzmann theory for a system size of $L=64$ at
intermediate densities is about 5 or 6, while in actual simulations
such number is smaller (1, 2 or 3 bands, but never more). This result
could be explained by the fact that the higher the number of bands the
more difficult to see them in simulations, because of band coalescence
at their early stages of formation. On the other hand, the Boltzmann
approximation was shown to be worse the lower $\gamma$, because
it predicted the jamming transition for $n=1/2$, a result that can be
valid for $\gamma$ close to $1/2$ (see Fig.\ \ref{phase}), but
certainly is not for small $\gamma$, where experimentally one finds
that it can be as low as $n\simeq 0.24$.

At both edges of the plot presented in Fig.\ \ref{figlam} the quantity
$\lambda_m$ goes to infinity as $\lambda_m\sim(n-1/2)^{-1/2}$ for
$n\gtrsim 0.5$ while $\lambda_m\sim(1-n)^{-1/4}$ when $n\lesssim 1$. The
former suggests that just after the transition only one band is formed;
the latter means that all bands coalesce into one which fills up the
whole system. As $\lambda_m\to\infty$, $k_m\to 0$; thus, as the
eigenvalues obey the restriction (\ref{k0relation}), the onset time of
the instability (given by $[{\rm Re}\;z({\bf k}_m)]^{-1}$) diverges
just above the transition. This is indeed observed in the numerical
simulations, where extremely long relaxation times are needed near the
jamming transition. For $\gamma$'s close to $1/2$ the onset time of the
instability also diverges (see Eq.\ (\ref{joseigen})). For intermediate
$n$ (in the jammed phase) the typical onset times range from a few
hundreds to a few thousands time steps.

The last remark of this subsection concerns the special case
$\gamma=1/2$. For this value, for which there is no distinction
between cars of type $\mu$ and $\nu$, Eq.\ (\ref{joseigen}) predicts
negative eigenvalues for any value of the density. As a consequence,
there is no jamming transition for this particular case, in agreement
with the numerical simulations (Fig.\ \ref{phase}a).

\subsection{Simulation of the Boltzmann Equations}
\label{BS}

All the information presented in the previous subsection was obtained
within the Boltzmann approximation, but in the linear regime. This
regime is only valid for small perturbations around the homogeneous
state. However when the perturbations are not longer small, as it
happens once the jammed phase appears, the full nonlinear Boltzmann
equation takes over, and the information given by the linear theory
may not hold. In particular, no linear theory can give the saturation
(i.e. completely filled nodes) occurring in the bands in the jammed
phase.

In order to  study the jammed phase we have performed simulations of
the full Boltzmann equations (\ref{Boltzeq}), starting with a uniform
state slightly perturbed at random. The evolution equations
(\ref{Boltzeq}) are then iterated until a stationary state is reached.
This final state can be either again uniform (when it is stable) or
non-uniform, exhibiting a traffic jam. This latter situation is
illustrated by Fig.\ \ref{phase4} for $n=0.6$ and $\gamma=0.2$, where
a plot of the average occupation profiles in the steady state is shown
on the line perpendicular to the bands. We clearly see the band
structure, with a completely saturated region ($u=1$ or $w=1$), bounded
by partly filled, very narrow layers. The result is similar to that
obtained in direct simulations on the microscopic model \cite{nos1}.
The center-to-center band distance, as measured from the simulations is
equal to 17 (in lattice units), while the linear stability analysis
predicts $\lambda_m\simeq 16$, in very good agreement.

The phase transition is numerically obtained for the same region of
parameters as those predicted by the linear theory ($n=1/2$ for all
$\gamma$). This result supports the idea that the fact that the
transition is $\gamma$-independent is an effect of the Boltzmann
approximation itself, and not of the linearization. In fact, from the
general theory of stability of dymanical systems, one can prove that
the stability is given by the linearized equations (unless
zero eigenvalues are obtained), not being modified by the
nonlinearities. Therefore it is the breakup of the correlations what
gives the degeneracy in the transition density. So, finding a
$\gamma$-dependent transition requires going beyond Boltzmann.
Some work along this line is currently being done.

{}From the simulation of the full Boltzmann equations we can obtain a
quantitative prediction for the velocity in the jammed phase. In order
to measure the velocity we cannot use the microscopic definition given
in Eq.\ (\ref{mvel}), because in the steady state the average
occupations do not change in time. However, we can use the modified
Eq.\ (\ref{corrt}), where no time variations appear, to compute the
velocity. Figure \ref{phase5} summarizes these results. In spite of the
$\gamma$-independent transition, there is a strong similarity between
this figure and the phase diagram obtained from the simulations of the
microscopic model (Fig.\ \ref{phase}); in particular, the slope $-1/2$
before the transition is properly obtained, as well as the behavior of
the remnant velocities after the transition has taken place.

In summary the Boltzmann approximation gives good predictions for many
physical quantities of the model, as remnant velocity, size of the
bands, onset time of instabilities, and predicts the existence
of a jamming transition, although it does not capture the fact that
the transition is $\gamma$-dependent. But with this simple scheme one
can obtain a qualitatively correct picture of the behaviour of the
system.

\section{A phenomenological model}
\label{sec:phenomeno}

We have seen in Secs.\ \ref{sec:micros} and \ref{sec:Boltzmann},
respectively, an exact microscopic and an approximate Boltzmann
descriptions of the model. Both of them consider
space-time variables that are discrete. In this section we are going
to present a continuous phenomelogical approach. Our motivation is
twofold. In the first place it provides a new theoretical
tool to understand the dynamics of the system. In the
second place it  makes a connection with the most usual approaches to
traffic flow problems: continuous fluid dynamics \cite{oldfluid}.
This section only tries to fix the main lines of a more complete
and exhaustive study of this traffic model from the point of view of
continuous dynamics that will be presented elsewhere.

\subsection{Equations}
\label{subsec-eqnmpm}

To define the continuous model we have to consider first the important
scales in the system, then redefine the variables, and finally take
the appropiate limit. Let us assume that the distance between city
crossings is $\epsilon$, that the linear size of the system is
${\cal L}\equiv L \epsilon$ and that there is  a scale of speed for
the cars $c$, defined as the speed of a car that moves freely (that is,
the length it will travel, if it does not find any red light or other
cars, divided by the spent time).
The time interval taken by half a cycle of a traffic
light (one time step) is then $\epsilon/c$.

Let us define a new set of space-time variables: $x\equiv n_x
\delta$, $y\equiv n_y \delta$ and $\tau \equiv n_\tau \delta/c$,
where $ \delta\equiv l \epsilon \ll 1 $
is a coarse grain length scale ($l$ is an integer such that
$1 \ll l \ll L$) we will define shortly, and $n_x, n_y, n_\tau$ are
positive integers in the interval $[0, L/l]$. Now we introduce, as
the variables of our model,
two density functions, one for each type of car: $u(x,y;\tau)$ and
$w(x,y;\tau)$, representing block averages in space and time of the
boolean variables $\mu_{\bf r}^t$ and $\nu_{\bf r}^t$:
\begin{eqnarray}
u(x,y;\tau) & \equiv& \frac{1}{l^3} \sum_{({\bf r},t)\in C_{(x,y)}
\times I_\tau} \mu_{\bf r}^t, \\
w(x,y;\tau) & \equiv& \frac{1}{l^3} \sum_{({\bf r},t)\in
C_{(x,y)}\times I_\tau} \nu_{\bf r}^t,
\end{eqnarray}
where $C_{(x,y)}$ is a square of size $\delta \times \delta$ ($l^2$)
sites, centered at
$(x,y)\equiv (n_x \delta,n_y \delta)$ and $I_\tau$ is a time
interval consisting of $l$ traffic light half-cycles (i.e. it lasts
a time $\delta_t  \equiv \delta/c$).
Notice that by definition the densities $u$ and $w$ are positive and
less than one:
\begin{equation}
 0 \leq u(x,y;\tau) \; ,\; w(x,y;\tau) \leq 1 \label{pos},
\end{equation}
and the dynamics, through the impossibility of having two cars at the
same site, imposses:
\begin{equation}
S(x,y;\tau) \; \leq \; 1, \label{exc_vol}
\end{equation}
where we have defined the new variable $S(x,y;\tau) \equiv
u(x,y;\tau)+w(x,y;\tau)$,
that stands for the total density of cars.

Now let us take the continuous limit by making $n_x, n_y, n_\tau
\rightarrow \infty$, $\epsilon \rightarrow 0$ and $l, L \rightarrow
\infty$ while keeping $\delta\rightarrow 0$ and $(x, y)\in\Omega$,
$\tau\in{\cal R}^+$, and ${\cal L}$ finite, where:
\[
\Omega =\{ (x,y): 0 < x, y < {\cal L}\}\in {\cal R}^2.
\]
In this way the functions $u(x,y;\tau)$ and $w(x,y;\tau)$ represent
coarse-grained local densities of the two types of cars. Because of
this definition these density functions lose track of the microscopic
details of the model (at scales $\epsilon$), and they are
supposed to be almost constant up to scales of order $\delta$.

Let us now write the equations that govern the evolution of
$u(x,y;\tau)$ and $w(x,y;\tau)$. We do so by writing a flux-balance
equation for the cars of each type going in and out of a block
$C_{(x,y)}$, that contains $l^2 \gg 1$ sites, with densities (constant
by definition) $u(x,y;\tau)$ and $w(x,y;\tau)$. The cars going in are
located either in the block $C_{(x-\delta,y)}$, inmediately to the
left, or in the block $C_{(x,y-\delta)}$ inmediately below. The cars
going out will go to the blocks $C_{(x+\delta,y)}$, located to the
right, or to the one above, $C_{(x,y+\delta)}$. Let us estimate how
many cars enter $C_{(x,y)}$ from $C_{(x-\delta,y)}$ in a time interval
$\delta_t$. In the first traffic light half-cycle only cars in
the column sited at the the right border of $C_{(x-\delta,y)}$ have
any chance to go into $C_{(x,y)}$. In average, the number of cars of
both types on that column are $u(x-\delta,y;\tau) l$ and
$w(x-\delta,y;\tau)l$. These cars have the chance to cross the left
border of $C_{(x,y)}$ and sit in its leftmost column. This column is
occupied, in average, by $S(x,y;\tau)l$ cars, leaving
$[1-S(x,y;\tau)]l$ empty sites. The number of cars that actually
go through depends on the relative positions of the cars in both
sides of the line and on the randomness built into the car movements.
For simplicity we assume that the distribution of cars on both sides
of the border is the most probable one: that in which the cars are
uniformly distributed along the line of length $\delta$. In
this case we can assume that in the first time step of $\delta_t$,
$\gamma u(x-\delta,y;\tau)[1-S(x,y;\tau)]l$ cars of type $\mu$ and
$\overline{\gamma} w(x-\delta,y;\tau)[1-S(x,y;\tau)]l$ of type $\nu$
will go from $C_{(x-\delta,y)}$ to $C_{(x,y)}$. During this
first time step there has been also movements of cars inside
$C_{(x-\delta,y)}$, but we are assuming that the densities
$u(x,y;\tau)$ and $w(x,y;\tau)$ are constant in space, throughout
the whole element, and in a time interval $\delta_t$. We
will see later that this assumption is consistent with the resulting
equations.  Under this condition we can suppose that the rightmost
column of $C_{(x-\delta,y)}$, and the leftmost column of $C_{(x,y)}$,
are always occupied by the same number of cars. Therefore the total
number of cars of type $u$ that will cross during $\delta_t$ ($l$
time steps) the left border of $C_{(x,y)}$ is:
\begin{equation}
N_{in \, \rightarrow}^u=
\, \gamma \, u(x-\delta,y;\tau) \, l
\,[1-S(x,y;\tau)] \, l. \label{3u}
\end{equation}
In the same way, the number of cars of type $u$ that will go inside
$C_{(x,y)}$ from the block below, $C_{(x,y-\delta)}$, is
\begin{equation}
N_{in \, \uparrow}^u =
\, \overline{\gamma} \, u(x,y-\delta;\tau) \, l \,
[1-S(x,y;\tau)] \, l, \label{1u}
\end{equation}
with the accompanying equations for $N_{in \, \rightarrow}^w$ and
$N_{in \, \uparrow}^w$ exchanging $u$ and $w$ and $\gamma$ and
$\overline{\gamma}$. Similarly we can obtain expressions for the
number of cars going out of $C_{(x,y)}$. Putting all together we can
write the balance equation for the increment in the number of cars of
each type in $C_{(x,y)}$:
\begin{eqnarray}
& & 2 l^2[u(x,y;\tau+\delta_t) - u(x,y;\tau)] = \nonumber \\
&= &N_{in \, \rightarrow}^u+ N_{in \,\uparrow}^u -
N_{ \rightarrow \, out}^u- N_{\uparrow \, out}^u \nonumber \\
&=&l^2 \, [\gamma \, u(x-\delta,y;\tau)  + \overline{\gamma} \,
u(x,y-\delta;\tau)][1-S(x,y;\tau)] - \nonumber \\
&-&l^2 \, u(x,y;\tau) \,  [1- \gamma \, S(x+\delta,y;\tau)-
\overline{\gamma} \, S(x,y+\delta;\tau)] \label{balance}
\end{eqnarray}
and the corresponding expression for $w$. The factor 2 comes from
considering that only cars from one of the sides can enter $C_{(x,y)}$
at each traffic light half-cycle. Observe that if we suppose that the
spatial variables in Eq.\ (\ref{Boltzeq}) are continuous,
we get exactly Eqs.\ (\ref{balance}). In this way we can think of
the coarse-grain treatment in this section as a continuous version of
the Bolztmann approximation of section \ref{sec:Boltzmann}.
Taylor expanding Eqs.\ (\ref{balance}) to first order in $\delta$ and
$\delta_t$ and fixing
$c=1$ to be consistent with the discrete model, we get the equation
for the time evolution of $u(x,y;\tau)$ and $w(x,y;\tau)$:
\begin{mathletters} \label{uwt}
\begin{eqnarray}
u_\tau & = & \frac{1}{2}\{\gamma\;[u(S-1)]_x +
\overline{\gamma}[u(S-1)]_y \} \\
w_\tau & = & \frac{1}{2}\{\overline{\gamma}[w(S-1)]_x +
\gamma[w(S-1)]_y \}
\end{eqnarray}
\end{mathletters}
\par \noindent where the subscripts $ \tau, x $ and $y$ mean,
respectively, $\partial /\partial \tau, \partial /\partial x$ and
$\partial /\partial y$. These equations have to be completed with the
periodic boundary conditions in ${\cal L}$ and initial conditions.

If Eqs.\ (\ref{uwt}) represent faithfully the original
microscopic discrete traffic flow model they should fulfill a few
neccesary (though, of course, not sufficient) conditions:
conservation of the number of cars, positivy of the densities,
excluded volume and the existence of stationary solutions describing
the two observed phases in the simulations. Let us check all of them.
First observe that Eqs.\ (\ref{uwt}) satisfy the following
conservation laws (constant number of cars of each type) for any
$\tau \in {\cal R}^+$:
\begin{equation}
\int_\Omega u(x,y;\tau) dx dy  \; = \;
\int_\Omega w(x,y;\tau) dx dy  \; = \; n {\cal L}^2/2 \, .
\end{equation}
Second, both densities of cars have to stay positive during the
evolution if the initial conditions were positive, and also
the total number of cars per site should be less than one (excluded
volume) (\ref{exc_vol}). It can be shown from a maximum principle
\cite{maximum} that both restrictions hold during the evolution
if the initial data $u(x,y;0)$ and $w(x,y;0)$ satisfy them. And
finally, from numerical simulations it is expected that the density
functions $u$ and $w$ corresponding to a stationary uniform state will
be solutions of equations (\ref{uwt}). It is also expected that a
function of band type (see Fig.\ \ref{phase}b), representing the jammed
state, will be a solution. It can be inmediately seen that the uniform
solution:
\begin{eqnarray}
u(x,y;\tau)& \; = \;& n/2, \nonumber \\ w(x,y;\tau) &\; = \;& n/2,
\label{homog}
\end{eqnarray}
is stationary.
It can also be checked that a jammed band configuration (of course
parallel multiple bands are also possible):
\begin{eqnarray}
u(x,y;\tau_0)&=&\theta (x-x_0-y) \theta(a+y-x+x_0), \nonumber \\
w(x,y;\tau_0)&=&\theta (x-x_0+a-y) \theta(y-x+x_0), \label{bandeqn}
\end{eqnarray}
where $\theta(x)$ is the Heaviside step function, is a stationary
solution too, but in the sense of the distributions
\cite{distribution}. Observe that the arguments of $\theta$ should be
taken ${\rm mod}\;{\cal L}$ due to the periodicity of the city. The
arbitrary parameter $x_0$ fixes the position of the band and $2 a=n
{\cal L}$ is its width.

\subsection{Stability}

We will proceed now to perform a stability analysis similar to
that in section \ref{LSA}. To study the linear stability of the
solution (\ref{homog}) let us perturb around the uniform solution:
\begin{mathletters}
\begin{eqnarray}
u(x,y;\tau) &=& n/2 + \delta u(x,y;\tau), \\
w(x,y;\tau) &=& n/2 + \delta w(x,y;\tau),
\end{eqnarray}
\end{mathletters} \par \noindent
where $\delta u,\delta w \ll n$. Substituing in (\ref{uwt}) we get,
to first order:
\begin{equation}
2 \, P _\tau \, = \, {\sf A} \, P _x \, +\, {\sf B}\,
P _y, \label{pt}
\end{equation}
where
\begin{equation}
P \equiv \left( \begin{array}{c} \delta u \\ \delta w \\
\end{array} \right)
\;\;\;\;\;
{\sf A} \equiv \left( \begin{array}{cc} \gamma
\left(\frac{3}{2}n-1\right)&\gamma \frac{n}{2} \\
\overline{\gamma} \frac{n}{2}& \overline{\gamma}
\left(\frac{3}{2}n-1\right)
\end{array} \right),
\end{equation}
and the expression for ${\sf B}$ is the same as the one for ${\sf A}$
exchanging $\gamma$ and $\overline{\gamma}$. Expanding in its Fourier
series, Eq.\ (\ref{pt})  can be easily integrated.
If we define the Fourier coefficients:
\begin{equation}
\widehat{P} (k_x,k_y;\tau) \; = \;
\int_\Omega \frac{dx dy}{{\cal L}^2} \, P (x_+,x_-;\tau) \, e^{-ik_x
x - i k_y y},
\end{equation}
with $k_{x(y)}=2 \pi n_{x(y)}/{\cal L}, n_{x(y)} \in {\cal Z}$,
(\ref{pt}) becomes:
\begin{equation}
2 \, \widehat{P} _\tau \,=\, ik_x {\sf  A} \,\widehat{P} \,+\,i k_y
 {\sf  B}\, \widehat{P}, \label{fpt}
\end{equation}
whose solution is:
\begin{equation}
\widehat{P} (k_x,k_y;\tau) \; = \;
e^{ i \frac{\tau}{2} (k_x {\sf  A} + k_y {\sf  B})} \; \widehat{P}
(k_x,k_y;0).
\end{equation}
The eigenvalues of $k_x {\sf  A} + k_y {\sf  B}$ are:
\begin{equation}
\lambda_{1,2}= k_+ \left( \frac{3}{2}n-1 \right )\pm
\sqrt{2 \zeta^2 k_-^2
(1-n) \left(\frac{1}{2}-n\right) +k_+^2 \frac{n^2}{4}}, \label{eigen}
\end{equation}
where $k_\pm \equiv (k_x \pm k_y)/2$.
The homogenous solution (\ref{homog}) will get unstable if the term
inside the square root in (\ref{eigen}) becomes negative, that is, if
\begin{equation}
2 \zeta^2 k_-^2
(1-n) \left( \frac{1}{2}-n \right) +k_+^2 \frac{n^2}{4}< 0
\label{cond}
\end{equation}
If $n>\frac{1}{2}$ condition (\ref{cond}) is always fulfilled, for any
value of $k_+$, for some big enough value of $k_-$. Once again we
obtain that the uniform solution becomes unstable for $n>\frac{1}{2}$.
Observe that Eq.\ (\ref{cond}) agrees to first order with the result
given in Eq.\ (\ref{joseigen}) when we make $(k_x,k_y)=(-k/\sqrt{2},
k/\sqrt{2})$. This is because Eq.\ (\ref{fpt}) is just the continuous
version of Eq.\ (\ref{linear}) to order $k$, showing again the
correspondence between both descriptions (see also the comments below
Eq.\ (\ref{balance})).

\subsection{Velocity}

Following the reasoning of Sec.\ \ref{subsec-eqnmpm} we can also get
a phenomenological estimate of the instantaneous average velocity in
this model.

The velocity is proportional to the number of movements that take
place in the system at any given time. We can measure the total
number of produced movements  by estimating how many cars cross
horizontal and vertical line elements of lenght
$\delta$.  For example the number of cars of type
$u$ and $w$ crossing a vertical line of length
$\delta$ centered at $(x,y)$ in a time step is (see Eq.\ (\ref{3u})
for a similar reasoning), respectively, $ u(x-\epsilon/2,y;\tau) l
[1-S(x+\epsilon/2,y;\tau)] \gamma$ and $w(x-\epsilon/2,y;\tau) l
[1-S(x+\epsilon/2,y;\tau)]\overline{\gamma}$ where we have explicitly
pointed out that the cars crossing the line are just {\it half} a
site to its left and they move to a place  half a  site to its right,
and therefore the corresponding densities should be
evaluated at $ x-\epsilon/2$ and $x+\epsilon/2$, respectively.
Similarly, the number of cars of type $u$ and $w$
crossing a horizontal line of length $\delta$ centered at $(x,y)$ is
(see (\ref{1u})), respectively, $u(x,y-\epsilon/2;\tau)l
[1-S(x,y+\epsilon/2;\tau)]\overline{\gamma}$ and $w(x,y-\epsilon/2;
\tau)l[1-S(x,y+\epsilon/2;\tau)]\gamma$. But when we take the limit
$\epsilon\to 0$ we can replace $u(x\pm \epsilon/2,y;\tau)$ by
\begin{equation}
u(x^\pm,y;\tau)\equiv \lim_{\epsilon\to 0^+}u(x\pm\epsilon/2,y;\tau),
\end{equation}
and $u(x,y\pm\epsilon/2;\tau)$ by
\begin{equation}
u(x,y^\pm;\tau)\equiv \lim_{\epsilon\to 0^+}u(x,y\pm\epsilon/2;\tau).
\end{equation}
Taking care of all the terms in the same way, properly normalizing
(the factor $1/2$ comes from the traffic lights) and taking the limit
$\epsilon \rightarrow 0$, $L \rightarrow \infty$, ${\cal L}$ finite,
we get the total speed of the system as:
\begin{eqnarray}
v&=&\frac{1}{2 n {\cal L}^2}\int_\Omega dx dy
\{ [1-S(x,y^+;\tau)] [ \overline{\gamma} u(x,y^-;\tau) +\gamma
w(x,y^-;\tau)] \nonumber \\ &+& [1-S(x^+,y;\tau)]
[ \gamma u(x^-,y;\tau)] + \overline{\gamma} w(x^-,y;\tau)  \}.
\end{eqnarray}
With this expression we can easily calculate the speed of the
homogenous solution (\ref{homog})
\begin{equation}
v=\frac{1-n}{2},
\end{equation}
in accord with all the previous models and the simulations.  We can
also calculate the speed for a solution of the jammed band type
(\ref{bandeqn}) getting, before taking the
limit inside the integral:
\begin{equation}
v=\frac{\gamma }{{\cal L} n} \epsilon = \lim_{L \rightarrow \infty}
\frac{\gamma }{ L n}.
\label{vband}
\end{equation}
When we take the limit $\epsilon \rightarrow 0$ (keeping ${\cal L}$
and $n$ finite) we get $v=0$. This result can be easily understood.
In one band the only possible movements are
along its border ($\sim \gamma L$) and
therefore the contribution to the velocity is of order $\gamma L / n
L^2$. This is exactly Eq.\ (\ref{vband}).
This makes us to conclude that  an infinite number
of bands is neccesary to have nonzero velocity in the jammed phase of
an infinite system. For example, if there is a band every $\lambda$
sites (see the discussion below Eq.\ (\ref{k0relation})),
 the velocity will be of order
\begin{equation}
\frac{\gamma}{n L} \frac{L}{\lambda} = \frac{\gamma}{n \lambda},
\end{equation}
which is finite in the $L\rightarrow \infty$ limit.

Let us return to Eq.\ (\ref{vband}). Though it is zero in the
$\epsilon \rightarrow 0$ limit it gives a prediction on the behavior
of the $v-n$ curves after the transition when $L$ is finite and
$\epsilon > 0$,
\begin{equation}
v=\frac{\gamma}{n L}.\label{vband2}
\end{equation}
To compare this prediction with the results of simulations we have
taken data from Fig.\ \ref{phase}. We have considered only the points
that represent stationary final states with one band \cite{nos1} and
we have plotted in Fig.\ \ref{1_over_n} the ratio $R\equiv L n v
/\gamma$ against $n$ for four different values of $\gamma$. If the
prediction of Eq.\ (\ref{vband2}) were completely correct the four
graphs should be the horizontal lines $R=1$. Instead, as can be seen
in the figure, the data fit well to four horizontal lines with
different values of $R$. The first important conclusion to draw from
Fig.\ \ref{1_over_n} is that, because the data fit to horizontal lines,
Eq.\ (\ref{vband2}) gives the correct behavior of the velocity after
the transition: decreases with $n$ as $1/n$ and is proportional to
$\gamma/L$. The discrepancy is in the proportionality factor, and may
be due to the fact that the jammed phase obtained in the simulations is
not a ``discontinuous" band (that goes sharply from one
to zero  in a distance $\epsilon$)
as the one in Eq.\ (\ref{bandeqn}). Observing carefully the
final configurations it can be seen that in fact the intermeditate
region where the density goes from one to zero, though narrow,
extends for several lattice spacings, being narrower the smaller is
$\gamma$ (the system is less random). The values of $R$ in Fig.\
\ref{1_over_n} confirm this behavior too.

\section{Conclusions}
\label{sec:conclusions}

In this paper we have developed a theoretical framework to describe
the two-dimensional cellular automata traffic flow model introduced
in Ref.\ \cite{nos1}. The starting point is the exact microscopic
evolution equations (\ref{eveq}) together with the microscopic
definition of the mean velocity (\ref{mvel}). From these basic
ingredients several approximate schemes can be carried out. The
simplest one is to follow the standard kinetic recipe of breaking up
spatial correlations into product of one particle averages. This
mean-field-like approach leads to a reasonably good description of the
model. In the first place, the $\gamma$-independent (for $\gamma >0$)
slope of the $v-n$ curves in the freely moving phase is correctly
obtained. In the second place, the existence of a phase transition
between the freely moving and the jammed states is also explained by
such a simple approximation. Nevertheless, it fails to describe
properly this transition because it predicts a $\gamma$-independent
transition density. Finally, this theory allows us to understand the
rich structure of the jammed phase. In this phase, the Boltzmann
approximation is able to explain the formation of bands completely
filled with cars, and the phenomenom of phase separation of the two
populations of cars inside these bands. Besides, it gives an answer to
the question posed by the simulations about the number of bands in the
jammed state. Our Boltzmann theory gives an estimation of the average
band-to-band distance. This distance appears to be finite for all
densities in the range $1/2<n<1$ (excluding the transition density
according to this approximation, $n=1/2$, and the full system $n=1$).
The conclusion is that there is an infinite number of bands in an
infinite system. On the other hand, the band-to-band distance
estimation let us evaluate the most probable number of bands for finite
$L$, once the values of $\gamma$ and $n$ are fixed. The obtained results
are in good agreement with the simulations.

In Sec.\ \ref{sec:phenomeno} we have faced the description of the model
with an utterly different phenomenological approach. This approach ends
up in a continuous model whose evolution is governed by a couple partial
differential equations. Although this approach seems to have no
relation to the Boltzmann approach, we have shown that there is indeed a
connection to the kinetic formalism -- from which it can be obtained
in the large spatial scale regime. The continuous approximation leads
to a connection between our model and those of fluid dynamics -- the
most usual tools to study one-dimensional traffic-flow problems. Apart
from its theoretical relevance, this phenomenological model is able to
provide a $\gamma$-dependent expression for the velocity as a function
of $n$ in the jammed phase. This result, together with the linearly
decreasing velocity obtained for the freely moving phase, completes a
theoretical description of the $v$-$n$ phase diagram.

The theory presented in this paper for a particular model opens new
ways in the theoretical study of two-dimensional traffic flow, which
deserve future development. First of all, it should be clear that
equations similar to (\ref{eveq}) can be written for any
cellular-automata-based traffic model and, therefore, the same
approximations can be applied to get information out of it. It
is also feasible to write down a continuous model, either by itself
or as a limit of the kinetic description. Secondly, it is possible to
devise another approximations based on the microscopic equations
(\ref{eveq}). For instance, it is our belief that the next step in
this scheme -- namely including two-point spatial correlations --
might account for the $\gamma$ dependence of the transition density.
We are presently working on it. Finally, it is worth to remark that
the approximate models reported in this work are far from being
exhausted. Questions such as the mathematical properties of the
instabilities in the continuous model or how to get an evolution
equation for the velocity remain unsolved.

\acknowledgements

We are indebted to Prof.\ J.B.\ Keller for a careful reading of the
manuscript and valuable suggestions. We also want to thank H.\
Bussemaker for his interesting comments on the Boltzmann
approximation, J.L.\ Vel\'{a}zquez, M.A.\ Herrero and A.\ Carpio
for helpful discussions concerning the continuous model, and A.\
S{\'a}nchez for discussions and collaboration in the early stages of
this work. Finally, we acknowledge financial support from the
Direcci\'{o}n General de Investigaci\'{o}n Cient\'{\i}fica y
T\'{e}cnica (Spain) through the projects PB92-0248 (F.C.M.\ and J.M.M.)
and PB91-0378 (J.A.C.\ and R.B.).

\noindent
\begin{figure}
\caption[]{Results of the simulations for the model of Ref.\
\cite{nos1} for a city with $L=64$: (a) phase diagram of the model
-- average velocity ($v$) vs.\  car density ($n$) --, exhibiting the
transitions from the  freely moving phase to the jammed phase for a set
of values of $\gamma$ (see \cite{nos1} for details); (b) snapshot of a
final run with $n=0.672$ and $\gamma=0.2$ (white and grey represent the two
types of cars, while black represents emty sites). Notice the phase
separation of the two populations of cars and the band-like traffic
jam.}
\label{phase}
\end{figure}

\begin{figure}
\caption[]{Plot of the real part of the largest eigenvalue of the
linear Boltzmann operator (only positive parts are seen). The set of
parameters is $n=0.8$ and $\gamma=0.1$. Regions where
${\rm Re}\; z({\bf k})$ is positive appear along the direction forming
an angle $\theta=3\pi/4$ with the $x$-axis (see text).}
\label{eigenrho=0.8p=0.1}
\end{figure}

\begin{figure}
\caption[]{Typical band-to-band distance, $\lambda_m$, in the jammed
phase, vs.\ the density, $n$, for several values of the randomness
$\gamma$.} \label{figlam}
\end{figure}

\begin{figure}
\caption[]{Jammed phase for the set of parameters $n=0.6$,
$\gamma=0.2$, as obtained from the simulation of the full Boltzmann
equations (\ref{Boltzeq}) on a lattice of size 64$\times$64. We only
plot here the occupation profiles along the line perpendicular to the
band. Cars of type $\mu$ ($\nu$) are represented by a dotted (dashed)
line. The total population if plotted with a solid line.}
\label{phase4}
\end{figure}

\begin{figure}
\caption[]{Average velocity of cars, $v$, as obtained from the
simulation of the non-linear Boltzmann equations (\ref{Boltzeq}),
vs.\ the car density, $n$. The transition to the jammed phase occurs at
$n=1/2$, regardless of the value of the turning parameter $\gamma$,
in contrast with the microscopic simulations (compare with Fig.\
\ref{phase}).}
\label{phase5}
\end{figure}

\begin{figure}
\caption[]{$R\equiv vLn/\gamma$ against the car density, $n$, for the
data taken from Fig.\ \ref{phase} in the single-band jammed phase (see
discussion after Eq.\ \ref{vband2}).}
\label{1_over_n}
\end{figure}


\begin{references}
\bibitem{note1} In principle, by virtue of the traffic lights -- which
introduce a time inhomogeneity, there might be two different steady
states: one for even and one for odd time steps. In that case, Eq.\
(\ref{avel}) would be the average of the mean velocity in those two
steady states.
\end{references}
\end{document}